\documentclass[revsymb, superscriptaddress]{revtex4}
\usepackage{amsmath,amsfonts,amssymb,graphicx}
\usepackage{float}
\floatstyle{plain}
\restylefloat{figure}
\restylefloat{table}

\begin{document}
\title{Discretising geometry and preserving topology I: \\ A discrete exterior calculus.}
\author{Vivien de Beauc\'e}
\affiliation{School of Mathematics, University of Dublin, Dublin 2, Ireland}
\author{Siddhartha Sen}
\email[e-mail:]{debeauce@maths.tcd.ie, sen@maths.tcd.ie}
\affiliation{School of Mathematics, University of Dublin, Dublin 2, Ireland}
\affiliation{IACS, Department of Theoretical Physics, Jadavpur, Kolkata, India}

\begin{abstract}
A discretisation scheme that preserves topological features of a physical problem is extended so that differential geometric structures can be approximated in a consistent way thus giving access to the study of physical systems which are not solely topological theories. Issues of convergence and a numerical implementation are discussed. The follow-up article covers the resulting discretisation scheme with metric data.
\end{abstract}
\maketitle

Problems in physics and engineering can often be formulated using the coordinate independent language of differential forms \cite{Flanders}. The setup can be characterized in terms of the following objects: a manifold $M$ on which $p$-forms (anti-symmetric tensor fields) $\omega_p$ are defined, the presence of a differential operator $d$ that maps a $p$-form to a $(p+1)$-form, a wedge product, $\wedge$, which allows a $p$-form and $q$-form to be multiplied to produce a $(p+q)$-form in a way that allows a Leibniz type rule to be valid for the operator $d$ when it acts on the product and the Hodge star operator $\star$. 

The Hodge star operator requires that the manifold $M$ have a metric. It maps a $p$-form to a $(D-p)$-form where $D$ is the dimension of the manifold $M$. Using the $\star$ operator a scalar product between $p$-forms can be introduced. This immediately allows the adjoint $d^{\dagger}$ of $d$ to be defined. The geometric discretisation scheme \cite{GD, adhep,  warner, hiptmair, dodzuik, albevario, starproduct} constructs an analogue model involving discrete objects for the continuum system described. Each of the elements ($\omega_p, \, d,\, \wedge, \, \star, \, d^{\dagger}$) now have discrete analogues. 

Two maps allow the approximation scheme to be defined precisely. The de Rham map that maps continuum variables, such as $\omega_p$, to discrete variables $\sigma_p$ ``chains'' and the Whitney map which maps discrete objects $\sigma_p$ to continuous variables $W[\sigma_p]$ ``cochains''. Results establishing the nature of the approximation and questions of convergence have been established \cite{dodzuik}. One generic feature of the scheme needs to be stressed. The discrete analogue of the Hodge star mapping $\sigma_p \rightarrow \star \sigma_p$ produces a different space of objects. Hence in order to properly work out discrete analogies of $d^{\dagger}$ it becomes necessary to work in the space which is symbolically $M_K \oplus \star M_K$ where $M_K$ is the discretisation of the manifold $M$. There is a doubling of spaces introduced \cite{GD}. 

In previous work it has been shown that the geometric discretisation scheme captured topological features of a system very nicely \cite{Us}. Extension of the ideas to accommodate fermions, in a Dirac-K\"ahler framework, have also been carried out \cite{viv,us}. Here we turn to the practical problem of geometrical information present in the original continuum problem in the discrete approximation scheme described without spoiling topological structures that it already encodes.
The approach we proceed to introduce, tackles the problem of defining what is known as a discrete exterior calculus and different attempts to tackle parts of this problem are available, Bossavit \cite{bossa} has introduced a discrete analogue of contraction by considering the dual mapping ``extrusion'' in the context of the electromagnetic theory. Hirani and co-workers \cite{hirani} have developed part of a discrete exterior calculus which is closely linked in spirit to the topological theory (the geometric discretisation described in I. being an example), but they do not study the Lie derivative which is of much importance here. 

In contrast to other works, in our approach the Whitney map which maps discrete objects to continuum variables plays an essential role. A definition of a product space of chains allow the notion of contraction to be defined which is essential for introducing analogues of the Lie derivative, an operation that describes the way objects change. 
The structure constructed gives rise to a consistent discrete exterior calculus, which is to our knowledge a new result. 
\section{The topological discretisation scheme}
The original geometric discretisation scheme captured topological data. For the simplical complex, it was developed by several authors \cite{GD, adhep,  warner, hiptmair, dodzuik, albevario, starproduct}. Here we will consider the hyper-cubic version of it \cite{samik} as our starting point. An alternative derivation of the Whitney elements is introduced which suits our purpose. First, we point out the important shortcomings of the original geometric discretisation method:
\begin{itemize}
\item
The discrete wedge was non-associative.
\item
Operators are only exact as operations on the discrete chains, (i.e there is no direct correspondence or ``functor'' between the discrete operations and the continuum ones).
\item
Whitney elements play a spectator role and so the theory is topological only. 
\end{itemize}
Recently, Samik Sen \cite{samik} has shown that by replacing a simplicial complex by a hyper-cubic complex, two new features emerge compared to the original proposal: the Hodge star operator satisfy $\star^K \star^L= Id$ and second, the associated dual complex being hyper-cubic as well, it can be provided with explicit expressions for the Whitney elements. 

The second item of the list above is crucial to us, because we are going to take the Whitney forms as the basic building blocks of the theory. Then, we use the product space of chains to generate the appropriate collection of forms which will in turn allow us to consider a discretised system with symplectic geometric data.

In contrast, the original scheme does not utilize the Whitney forms beyond providing a space that gives rise to the de Rham co-homology complex, so we need not know the explicit form of the Whitney elements to proceed topologically.

Here, we put the Whitney elements and the various combinatorial manipulations of them at the forefront of the formulation. Here, a discrete operation is {\it exact \it} if when mapped to the continuum its algebraic properties hold exactly. Then in turn, by accommodating the other operations in the theory ($\wedge, \, d, \, \star$) we can show that the combinatorics of Whitney elements that we introduce is such that the topological data is still captured. We find that:
\begin{itemize}
\item
The basis of Whitney forms encodes the topology and is very constraining that way, but allows for the product space of chains.
\item
Secondly, the problem of the non-associativity of the wedge is by-passed by not relying on the wedge but on a new product  in order to define the contraction.
\end{itemize}
These arguments both provide a strong motivation for considering the product space of chains. Before we turn to that, we review the geometric discretisation scheme.
\subsection{Differential forms, manifolds and complexes}
The general program of the topological discretisation is described at length in \cite{Us}, On a manifold, we have fields $\phi^{(p)}$ which are differential forms (antisymmetric tensor fields), and the various operations of interest are:
to consider a theory defined on an
arbitrary manifold M of dimension D. Suppose the theory has fields
$\phi^p(\vec{x})$, where $\vec{x}\in M$ and $p=0,1,\dots,D$.  The theory is constructed using the following
objects which are defined on the manifold M : p-forms $\phi^p$ which
are generalized antisymmetric tensor fields, the exterior derivative
$d:\phi^p \to \phi^{p+1}$, the
Hodge star operator $*:\phi^p \to \phi^{D-p}$ ,which is required to define 
scalar 
products, and the wedge operator $\phi^p \wedge \phi^q = \phi^{p+q}$. On a manifold M of dimension D, the operation ($\wedge, *, d$) on
p-forms, $\phi^p (p=0,\dots,D)$, satisfy the following: 
\begin{enumerate}
\item $\phi^p\wedge\phi^q = (-1)^{pq}\phi^q\wedge\phi^p$. 
\item $d(\phi^p\wedge\phi^q) = d\phi^p \wedge \phi^q + 
(-1)^p\phi^p\wedge d\phi^q$.
\item $*\phi^p = \phi^{D-p}$.
\item $* * = (-1)^{Dp+1}$.
\item $d^2=0$, $(d^*)^2=0$. 
\item $d^\dagger=(-1)^{D(p+1)+1}*d*$. ($d^\dagger$ is the adjoint of $d$).
\end{enumerate}
The following definitions are also important: 
\begin{itemize}
\item The Laplacian on p-forms $\Delta_p = d_{p-1} d_p^{*}+d_{p+1}^{*}d_p$. 
\item The inner product $(\phi^p,{\phi^p}')=\int_M\phi^p\wedge *{\phi^p}'$.
\end{itemize}
\noindent
We would like to discretise the fields of $\phi^p$, the inner product $(\cdot,\cdot)$ and the operators $\wedge, *, d$ such that discrete analogies of the above interrelationships hold. To do this it is necessary to first introduce a few standard ideas. We first discretise the manifold M by replacing M by a collection of discrete objects, known as simplices, glued
together \cite{NS}. For $p\geq 0$, a $p$-simplex $\sigma^{(p)}$ is defined to be the convex hull in some
Euclidean space $\mathbb{R}^m$ of a set of $p+1$ points $v_0,v_1,\dots,v_p\in
\mathbb{R}^m$
where the vertices are independent in that 
\[\sum_{i=0}^p\lambda_i v_i=0 \hspace{1 in}    \sum_{i=0}^p\lambda_i=0 \]
implies that $\lambda_i=0$ for $i=0,\dots ,p$ where 
$\lambda_0, \lambda_1,\dots ,\lambda_p$ are real numbers. Geometrically the {\em barycenter} of a given n-simplex $\sigma^{(n)}=
(v_0,v_1,\dots,v_n)$ is defined by
\[
\hat{\sigma}^{(n)}=\frac{1}{n+1}(v_0 +v_1+\dots+v_n).
\]

\noindent
Thus the barycenter of $\sigma^{(1)}=[v_0,v_1]$ is the 
midpoint between the vertices $v_0$ and $v_1$. We can now describe a particular way that a given manifold $M$ can be 
discretised. Let $S$ be a collection of simplices $\{\sigma^{(n)}_i\}$, 
$n=0,1,\dots,D$, with the property that the faces of the simplices which 
belong to $S$ also belong to it. The elements of $S$ glued together in 
the following way is known as a simplicial complex\cite{NS}, \cite{HY}: 
\begin{enumerate}
\item $\sigma_i^{(n)} \cap \sigma_j^{(k)} = 0$ if
$\sigma_i^{(n)}$, $\sigma_j^{(k)}$ have no common face. 
\item $\sigma_i^{(n)} \cap \sigma_j^{(k)} \neq 0$ if
$\sigma_i^{(n)}$, $\sigma_j^{(k)}$ have precisely one face in common,
along which they are glued together. 
\end{enumerate}
In many cases of interest(including all differentiable
manifolds \cite{HY}), $M$ can be replaced by a complex $K$ which it is
topologically equivalent to. $K$ is then said to be a triangulation of $M$(Note this 
triangulation is not unique).  \noindent
In this way of discretising M, the building blocks are zero, one, $\dots$, 
D-dimensional objects, all of which are simplices e.g.generalized 
oriented tetrahedra.

The same manifold can be discretised using hyper-cubes or a CW complex. We will adopt the hyper-cubic approach which is the neater one, for the following reason which has to do with the construction of the complex dual $L$. First note that in contrast with FIG.1, the analogue construction for the dual complex in the case of the hyper-cubic complex leads to two superimposed complexes shifted by a half edges length, and in turn the Whitney elements are readily available \cite{samik}. 

\subsection{The discrete operations.}
Let us first recall the construction of the discrete Hodge star for the simplicial complex which leads to $L$. In the way analogue to the continuum property (3. above), a map from a $(D-p)$ dimensional object to a $p$ dimensional object was introduced.
Now consider the barycentres
of the building blocks of the system $K$. 
We first construct $(D-p)$ dimensional objects whose vertices are barycentres of a sequence of successively higher
dimensional simplices, where each simplex is a face of the following one. In other words $(D-p)$ dimensional objects of the form 
$\{\hat{\sigma}_p, \hat{\sigma}_{p+1}, \dots, \hat{\sigma}_D\}$, 
where $\sigma_n$ is a face of $\sigma_{n+1}$. The orientation of these are set so as to be compatible with the manifold.
Joining these objects together gives us the dual of $\sigma_p$.

\begin{figure}
\begin{center}
\includegraphics[width=200pt]{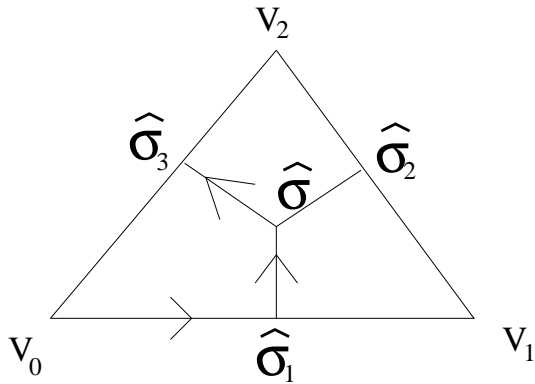}
\caption{Dual complex $L$.}
\end{center}
\end{figure}

We can now give the general rule for mapping an n-simplex
$\sigma_n = [v_0,\dots,v_n]$ to a ($D-n)$ dimensional object (
(D-n) cell) as follows:
We think of $\sigma_n$ as an element of a simplicial complex K. We have
\[*_K:[v_0,\dots,v_n]\to\cup[\hat{\sigma_n},
\hat{\sigma_{n+1}},\dots,\hat{\sigma_D}], \]
where $\hat{\sigma}_{n+1}$ is the barycenter of an ($n+1$)-simplex which 
has $\sigma_n$ as a face.
$\hat{\sigma}_{n+2}$ is the barycenter of an ($n+2$)-simplex 
which has $\sigma_{n+1}$ as a face and so on. These objects have to
be coherently oriented with respect to $[v_0,\dots,v_n]$. The set
of these cells constitutes the dual space $\hat{K}$ of K. By this procedure, a discrete version of the Hodge star operation was constructed. Let us explain. The Hodge star operator involves forms.
It maps $p$-forms in D dimensions to a ($D-p$)-form. The $*_K$ map 
involves not forms but geometrical objects. There is a 
simple correspondence relation between these two cases. 
Given a $p$-form, $\phi_p$, and a p-dimensional geometrical space, $\Sigma_p$, 
the $p$-form can be integrated over $\Sigma_p$ to give a number. 
Thus $\Sigma_p$ and $\phi_p$ are objects that can be paired.
We can write this as a pairing
\[ (\phi_p, \Sigma_p) =\int_{\Sigma_p}\phi_p .\]
In order to proceed, we need to introduce some more structure.
We start by associating with a simplicial complex $K$, containing
$\{\sigma_p^i\}$ $(i=1,\dots,K_p;p=0,\dots,D)$ a vector space
consisting of finite linear combinations over the reals of the
$p$-simplices it contains. This vector space is known as the space of
p-chains,$C_p(K)$.
For two elements $\sigma_p^i, \sigma_p^j \in C_p(K)$, a scalar
product $(\sigma_p^i,\sigma_p^j)=\delta_j^i$ can be introduced.
An oriented p-simplex changes sign under a change of orientation i.e.
if $\sigma_p=[v_0,\dots,v_p]$ and $\tau$ is a permutation of the 
indices $[0,\dots,p]$, then $[v_{\tau(0)},\dots,v_{\tau(p)}]=
(-1)^\tau[v_0,\dots,v_p]$, with $\tau$ denoting the number of transpositions
needed to bring $[v_{\tau(0)},\dots,v_{\tau(p)}]$ to the order
 $[v_0,\dots,v_p]$.

Given the vector space $C_p(K)$, the boundary operator $\partial^K$ can
be defined as
\[ \partial^K:C_p(K)\to C_{p-1}(K). \]
It is the linear operator which maps an
oriented p simplex $\sigma^{(p)}$ to the sum of its (p-1) faces
with orientation induced by the orientation of $\sigma^p$.
If $\sigma^p=[v_0,\dots,v_p]$, then 
\[ \partial\sigma^p=\sum_{i=0}^p (-1)^i[v_o,\dots,\hat{v_i},\dots v_p],\]
where $[v_o,\dots,\hat{v_i},\dots, v_p]$ means that the vertex $v_i$
has been omitted from $\sigma^p$ to produce the face
``opposite'' to it.

Given that $C_p(K)$ is a vector space, it is possible to define
a dual vector space $C^p(K)$, consisting of 
dual objects known as cochains;  that is we can take an element
of $C_p(K)$ and an element $C^p(K)$ to form a real number. 
Since the space $C_p(K)$ has
a scalar product, namely if $\sigma_p^i, \sigma_p^j \in C_p(K)$
then $(\sigma_p^i,\sigma_p^j)=\delta_{ij}$. 
We can use the scalar product to identify $C_p(K)\equiv C^p(K)$,
so 
that we can consider oriented $p$-simplices as elements of $C^p(K)$ as well
as $C_p(K)$. We can write our boundary operation as
\[ ([v_0,\dots,\hat{v_i},\dots,v_p],\partial^K[v_0,\dots,v_p])=(-1)^i.\]
This suggests introducing the adjoint operation $d_K$ defined as
\[
(d^K[v_0,\dots,\hat{v_i},\dots,v_p],[v_0,\dots,v_p])=
([v_0,\dots,\hat{v_i},\dots,v_p],\partial^K[v_0,\dots,v_p]).
 \]
This is the coboundary operator which maps $C_p(K)\to C_{p+1}(K)$. 

Indeed we have
\[ d^K[v_0,\dots,v_p]=\sum_v[v,v_0,\dots,v_p], \]
where the sum is over all vertices $v$ such that
$[v,v_0,\dots,v_p]$ is a (p+1) simplex.

The boundary operators $\partial_K$ and the coboundary
operator $d_K$ have the property $\partial_K\partial_K=d_Kd_K=0$.
Furthermore,
\begin{align*}
d_K:C_p\to C_{p+1}, \\
 \partial_K:C_p\to C_{p-1}.
\end{align*}
These operators are the discrete analogies of the operators
$d$ and $(-1)^{D(p+1)+1} *d*=d^\dagger$ which act on forms.

These operators could be defined only when a scalar
product(``metric'') was introduced in the vector space $C_p$'s.
At this stage we have a discrete geometrical analogue of 
$d$, $d^\dagger$ and $\star$. We have also commented on the fact that the operation
$\star$ maps simplices into dual cells i.e. not simplices.  If the original 
simplicial system is described in terms of the union of the vector spaces of all
p-chains then the space into which elements of the vector space
are mapped by $\star$ is not contained within this space, unlike the
situation for the Hodge star operation on forms. It is found that this leads inevitably to a doubling of the fields when
discretisation, preserving topological structures, is attempted.
\subsection{Relation to the Whitney map}
We now need a way to relate a $p$-chain to a $p$-form. This 
together with a construction which linearly maps p-forms to p-simplices
allows us to translate expressions in the continuum to a corresponding
discrete geometrical objects. We start with the construction of the linear
maps from p-chains to $p$-forms due to Whitney \cite{Whitney}. In order to define this map, we need to introduce barycentric coordinates associated with a given p-simplex $\sigma^p$. Regarding $\sigma^p$ as an element of some $\mathbb{R}^N$, we introduce
a set of real numbers $(\mu_0,\dots,\mu_p)$ with the property
\begin{align*}
\mu_i \geq 0, \\
\sum_i \mu_i=1.
\end{align*}
A point $x\in \sigma^p$ can be written in terms of the vertices of 
$\sigma^p$ and these real numbers as
\[ x=\sum_{i=0}^p \mu_i v_i. \]
Note if any set of $\mu_i=0$ then the vector $x$ lies on a face of $\sigma^p$. One can think of $x$ as the position of the
center of mass of a collection of masses $(\mu_0,\dots,\mu_p)$ located on the vertices $(v_0.\dots,v_p)$ respectively. Setting $\mu_i=0$ for instance means the center of mass will be 
in the face opposite the vertex $v_i$. The Whitney map can now
be defined. We have
\[ W^K:C^p(K)\to\Phi^p(K), \]
where $\Phi^p(K)$ is a p-form. If $\sigma^p\in C^p(K)$ then
\[ W[\sigma^p]=p!\sum^p_{i=0}(-1)^i\mu_i d\mu_0\wedge\dots
\hat{d\mu_i}\wedge\dots d\mu_p, \]
where $\hat{d\mu_i}$ means this term is missing, and $(\mu_0,\dots,
\mu_p)$ are the barycentric coordinate functions of $\sigma^p$.

We next construct the linear map from $p$-forms to $p$-chains. This is 
known as the de Rham map. We have
\[ A^K:\Phi^p(K)\to C^p(K), \]
defined by
\[ <A^K(\Phi^p),\sigma^p> = \int_{\sigma^p}\Phi^p, \]
for each oriented $p$-simplex $\in K$. 

%
\noindent
A discrete version of the wedge product can also be 
defined using the Whitney and de Rham maps such that 
$\wedge^K:C^p(K)\times C^q(K)\to C^{p+q}(K)$ as follows:
\[ x \wedge^K y = A^K(W^K(x)\wedge W^K(y)).\]
It has many of the properties of the continuous wedge product in that 
it is skew symmetric and obeys the Leibniz rule but it is non-associative.

At this stage we have introduced all the building blocks necessary
to discretise a system preserving topological structures. We summarize
the properties of the maps introduced in the form of a theorem. We have 
\noindent

{\bf Theorem(Whitney\cite{Whitney})}
\begin{enumerate}
\item $A^K W^K$ = Identity. \\
\item $dW^K = W^K d^K$, where $ d:\phi^p \to \phi^{p+1}$. \\
\item $\int_{\mid\beta\mid}W^K(\alpha)=<\alpha,\beta>$, $\alpha, \beta
\in K$\\
\item $ d^K A^K = A^K d $. \\
\end{enumerate}

\noindent
This theorem shows how $d^K$ can be considered as the discrete analogue of
$d$. We now show how $*^K$ can be considered as discrete analogue of
$*$. For this we need barycentric subdivision.

In order to construct 
the star map, two geometrically distinct spaces were introduced. The 
original simplicial decomposition $K$ with its associated set of p-chains
$C^p(K)$ and the dual cell decomposition $L$ with its associated 
set of p-chains $C^p(L)$. 
These spaces are distinct. However both belong to the
first barycentric subdivision of $K$. This allows the use the $*^K$
operation if we think of $K$ and $L$ as elements of $BK$.

We proceed as follows. Let BK and $L$
denote the barycentric subdivision and dual triangulation, and
\[ *^K:C^p(K)\to C^{n-p}(L). \]
However $C^p(K)$ and $C^p(L)$ are both contained in 
$C^p(BK)$ as we have seen. Let 
\[ W^{BK}:C^p(BK)\to\phi^p(M),\]
denote the Whitney map. Then we have \cite{adhep, samik}, for $x\in C^p(K), y \in C^{n-p-1}(L)$,
\begin{enumerate}
\item $<*^K x,y > = \int_M W^{BK}(Bx)\wedge W^{BK}(By).
\\
<*^{L} y,x > = \int_M 
W^{BK}(By)\wedge W^{BK}(Bx). $
\item $\partial^K = (-1)^{np+1}*^{L}d^{L}*^K$ on $C^p(K)$.
$\partial^{L} = (-1)^{nq+1}*^Kd^K*^{L}$ on $C^q(L)$.
\end{enumerate}
for the hyper-cubic complex.
\noindent
These are the discrete analogies of the interrelationships between $d, d^\dagger,
*$ and $<\cdot,\cdot>$ in the continuum.

Note $K\neq L$ and that properties of $\partial_K, d_K$
analogous to those for differential forms only hold if 
$K$ , $L$ are both regarded as elements of BK. 
This feature of the discretisation method is, as we shall see, crucial 
if we want to preserve topological properties of the original system. 

In the next section we derive a Whitney map in the context of a hyper-cubic complex (which matches the one given in \cite{samik}). In section III, the main features of the new scheme are introduced in relation with the definition of a contraction operation; in section IV, we proceed with the Lie derivative, the Lie algebra and associated group action on the space of Whitney forms. In the last part we discuss the convergence and a numerical implementation. In the Appendix we outline in what sense the theory still captures topology, and display some calculations.
The subsequent article will tackle the metric along with the covariant derivative and the curvature. A special role will be given to the dual complex $L$ in relation with the metric connection. We will also propose some applications to physics.

\section{The hyper-cubic complex}
In what follows, we will consider a hyper-cubic complex, since it is the most used in the literature, and we have already pointed out its advantages. We consider singular homology rather than homology (see the definition of a simplicial complex \cite{NS}). Let us describe how to handle the new setting. Consider the surface of the cube, topologically the two-sphere $S^{2}$. Consider the face $[0123]$, and let us follow the steps of the construction of the Whitney map given in the last section. The zero-form Whitney elements are obtained by application of the boundary conditions and are given by:
\begin{equation}
\{\mu_{0} = (1-x)(1-y), \;
\mu_{1} = x (1-y), \;
\mu_{2} = xy, \;
\mu_{3} = y (1-x)\}.
\end{equation}

\begin{figure}[H]
\label{fig:T}
{\par\centering \includegraphics{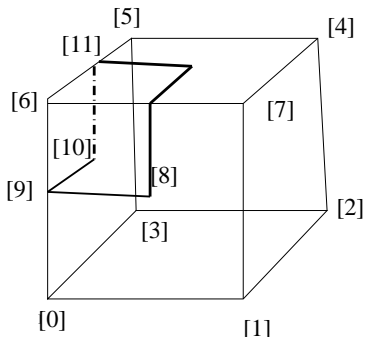} \par}
\caption{The cube with labels used throughout the discussion. The thick loop is a holonomy around the vertex $[6]$, discussed in the follow-up article.}
\end{figure}

The book-keeping for the co-boundary operator is as follows. Consider the simplex $[0123]$. Then the application of $d_{K}$ is shown below. 
\begin{align}
d^K [0] &= [10] + [30] \\
d^K [01] &= [0123] \\
d^{K} [10] &= [1023] = - [0123] \\
d^{K} [0123] &= 0
\end{align}
Besides, the orientation is fixed by specifying a two-chain in this case $[0123]$.
The way to check that the $d$ operation is correct, is by verifying that the Stokes theorem holds. However, by using the formulae of Whitney directly given the cubic barycenter coordinates one finds that the requirement $W^K d^K= d W^K$ is not satisfied which then makes the basis list of forms an incorrect basis of Whitney elements as noted by Samik Sen \cite{samik}. 

To remedy this problem we derive the Whitney map by induction. A Whitney map is defined as follows. Take the barycenter coordinates satisfying
\begin{equation}
\mu_i([j]) = \delta_{ij}
\end{equation} 
and generate the Whitney elements by application of
\begin{align}
\label{didab}
d W^{(r)} &= W^{(r+1)} d^K, \\
\label{bc}
< \sigma^{(r)}_i,\, W^{(r)}(\sigma^{(r)}_j)> &= \delta_{ij}.
\end{align}
which is used as the definition of $W^{(r+1)}$ given $W^{(r)}$. The Whitney elements thus obtained satisfy automatically the Stokes theorem.  \\



\noindent
The reader may check that the list of all Whitney elements given in the appendix A do indeed satisfy this rule (and may skip the rest of this section).

We will sketch the proof by induction, a detailed one will be given elsewhere. The first observation is that for zero co-chains, adopting the functional form of the example of the square above, we find that any vertex $[i]$ has within a given top dimensional simplex $\triangle^n$ of which it is a face the following expression:
\begin{equation}
W ([i] ) = \Pi_k f_k
\end{equation}
where each factor is either
\begin{align}
f_i &= f^{+}_i =(1-x_i) \\
&or \\
&= f^{-}_i =  x_i. 
\end{align}
The factors enforce the boundary conditions \eqref{bc}, and depend on which bounding simplex we are taking for the evaluation of the Whitney map. The claim is then that we can extract the remaining elements, i.e of higher degrees. An immediate observation is that application of $d$ will suppress factors $f_i$ and replace them up to sign by a form $dx^i$. Moreover, the factor $f_i$ that is suppressed relaxes one of the boundary conditions relating to the simplex containing both the original one and its extension caused by application of $d^K$. In turn, this leads to a well-defined pull-back map for integration over a given simplex $\sigma^{(r)}_{i}$ which we denote by
\begin{equation}
< \sigma^{(r)}_i, \; . \; >^\star: \Omega^r(M, \, \mathbb{R}) \longleftarrow \mathbb{R} 
\end{equation} 
and the space of Whitney elements is generated by the image (denoted $\Im$) of $1 \in \mathbb{R}$ under the pull-back for each and every $r$-simplex in $\sigma^{(r)}_i \in \triangle^{(r)}(K, \, \mathbb{R})$,
\begin{equation}
W^{(r)} \triangle^{(r)} (K, \, \mathbb{R}) = \Im (<\triangle^r(K),\; . \;>^\star 1).
\end{equation}
 This map, gives us the Whitney elements. Take a $\sigma^{(r)}_i$, then assume $W^{(r)} (\sigma^{(r)}_i)$ is given. Then by application of the pull-back map after evaluating $dW(\sigma^{(r)}_i)$ using \eqref{didab} gives $W^{(r+1)}$ for a $(r+1)$-simplex. This process is done until the list of $(r+1)$-simplices is exhausted. 

Consider a basic example; in the appendix take $\mu_0$. Then apply the exterior derivative to get a linear combination of terms which match $W([01])$, $W([03])$ in the square $[0123]$. Then select $W([01])$.

A side issue is that of completeness. One might be under the impression that we are calculating $W$ for exact forms in 
\begin{equation}
Z^r(K, \, \mathbb{R})
\end{equation}
Obviously, this is not true, the pullback allows us to evaluate the Whitney map on the basis set of $C^r(K, \, \mathbb{R})$ which is the collection of simplices $\triangle^r(K)$. The Whitney elements obtained are the correct ones (see the appendix). 

To visualize the Whitney elements, we sketch below what a 1-co-chain looks like. As is apparent from the formulae, a one-chain with non-zero coefficients along the edges parallel to the $y$-axis have the following shape, you see that the form jumps over cells which are parallel to the 1-chain but this does not affect the differentiability of the form since $d = dx \wedge \frac{\partial}{\partial x}$ on a $dy$-form. and in multiple dimensions you get all derivatives apart from $\frac{\partial}{\partial y}$.
\begin{figure}[H]
\label{fig:T}
{\par\centering \includegraphics{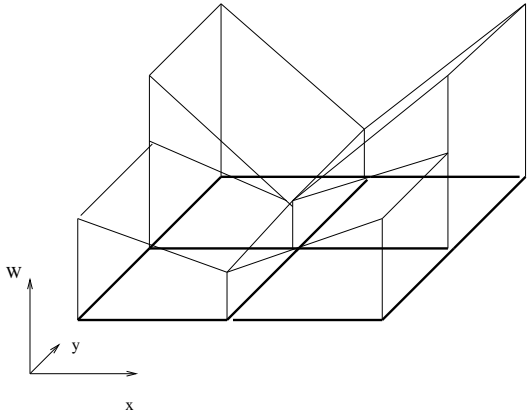} \par}
\caption{The two dimensional plane with a square complex, coordinates $(x,\, y)$ and a Whitney form with function coefficient sketched in the $W$-axis.}
\end{figure}
At this point we take the topological theory above and modify it in order to consider geometry. The key construction that leads us to geometry is the formulae given below for a hyper-cubic interior product and the numerous properties it possesses as we proceed to describe.


\section{Setting up the exterior calculus}
In order to extend the range of applicability of the method we need to be able to study ``flows''. This involves introducing the idea of a vector field in the discrete setting and constructing an analogue of the Lie derivative. To do this we need to define an discrete analogue of the interior product. Geometrical information in the sense of introducing a metric in this frame corresponds to following the vielbein formalism and is related to the Regge approach. We postpone consideration of this important construction to the follow-up article and focus on vector fields and the Lie derivative and algebra. The metric is the locally flat metric mentioned in section I.  

We now begin our construction by noting that the discretisation scheme is too restrictive in order to consider a geometry. What we need is a contraction map $i_v$ in order to consider vector fields and geometry. With $d$ and $i_v$ we will construct the Lie derivative $L_v$ (and in turn, in the subsequent article, with some prescription for the metric and the Lie derivative we will be able to construct a covariant derivative $\nabla_v$). 
\subsection{Interior product}
The problem we start with is to construct a hyper-cubic analogue of the interior product (or contraction) map:
\begin{equation}
\label{thecontra}
i_v \omega^{(r)} = \frac{1}{r!} v^{\nu} \omega_{\nu \mu_2 \ldots \mu_r} dx^{\mu_2} \wedge \ldots \wedge dx^{\mu_r}
\end{equation}
Our strategy is to apply the de Rham map to the dual one form of the vector field $v$ represented as a one-chain (i.e a linear combination of edges) and to the degree $(r-1)$-form that results from $\omega$ after contraction. Thus providing us with a pair of chains. Let us extract it from the continuum by rewriting \eqref{thecontra} as
\begin{equation}
\label{thecontrac}
i_v \omega^{(r)} = \frac{1}{r!}  v^{\nu} \, \rfloor^K  \omega_{\nu \mu_2 \ldots \mu_r} dx^{\mu_2} \wedge \ldots \wedge dx^{\mu_r}, 
\end{equation}
and then notice that this expression, as a pairing of two forms, can be represented on the chain space of the $K$ complex as:
\begin{equation}
i_{\alpha^{(1)}} \sigma^{(r)} = \alpha^{(1)} \rfloor^K \eta^{(r-1)}(\sigma).
\end{equation}
where $\alpha$ is a one-chain, $\sigma$ is an $r$-chain and $\eta$ an $(r-1)$ chain which results from contracting $\sigma$. It is similar to what we have in the continuum with exception that $\alpha$, a one-chain plays the role of a coefficient function and so should be treated as such. Second remark is that we will see that in fact this leads to a well-defined operation on the complex and is exact when the form $\omega$ and the dual to the vector field $v$ are in fact Whitney elements. Also, we introduced a generalized product on the space of chains, denoted by $\rfloor^K$ the choice of the symbol is made to stress that it comes from the contraction operation. It clearly distinguishes between the chain on its right and that on its left. Let us postpone the formal construction and consider examples. Start with a $1$ D example. Consider the edge $[01]$ on the line. Then, the Whitney elements are
\begin{equation}
\{ W([0]) = 1-x, \, W([1]) = x , \, W([01]) = dx.\}
\end{equation}
and
\begin{equation}
i_{[01]} [01] = [01] \rfloor^K ([0] + [1]). 
\end{equation}
That is, we took the sum of the two vertices that remain when we remove the edge $[01]$. Also, the construction is guided by the embedding in the continuum:
\begin{equation}
W(i_{[01]} [01]) = W([01]) \times W([0] + [1]) = dx \times 1 = 1. 
\end{equation}
Again we said that the one chain $\alpha$ plays the role of a function coefficient so, in this case it is $1$ and hence the result. If you contract $[0]$ or $[1]$ with the edge $[01]$, you get zero because the edge $[01]$ is not a face of either vertices. Consider the 2D example of the square.
\begin{figure}[H]
\label{fig:T}
{\par\centering \includegraphics{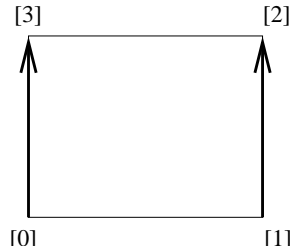} \par}
\caption{Two parallel edges in a given square which arise from contraction of $[0123]$ with one of the edges $[01]$ or $[32]$.}
\end{figure} 
 The Whitney elements are:
\begin{align}
\{&W[0] = (1-x)(1-y), \, W[1] = x(1-y), \, W[2] = xy, \, W[3] = (1-x) y,
\, W([01]) = (1-y) dx, \\
&W([12]) = xdy, \, W([32]) = ydx, \, W([03]) = (1-x )dy, \, W([0123]) = dx \wedge dy.\}
\end{align}
This time there are many combinations of two chains which lead to a non-zero contraction, for example the contraction of the square with an edge, say $[01]$. let us see two examples which will help us to specify $i_v$.
\begin{align}
i_{[01]} [0123] &= [01] \rfloor^K\, ([03] + [12]) \\
i_{[01]} [32] &= [01] \rfloor^K\, ([3] + [2]).
\end{align}
And they lead to
\begin{align}
W (i_{[01]} [0123]) &= W([01]) \rfloor^K W([03]+[12]) = (1-y) dx \rfloor^K\, dy = (1-y)dy \\
W(i_{[01]} [32])  &= W( [01]) \rfloor^K W([3] + [2]) = (1-y)dx \rfloor^K \, x = (1-y)x.
\end{align}
To comment, the first calculation shows how when contracting a 2-chain, we get a linear combination of the edges perpendicular to the edge which specifies the vector field. This generalizes obviously to higher degree chains: we get a linear combination of the $(r-1)$-simplices which are orthogonal to the vector field (an edge). The second calculation is there because it shows that although $[01]$ is not a face of $[32]$, the contraction still applies because they are parallel within the same cell of highest degree which plays the role of the local open set.

We formalize a little the interior product we have constructed and in the next sections, we will have to modify the other operators accordingly, notably $d^K$ the co-boundary.

The discretised interior product is a map from the product space of chains to the product space of chains, in the way that $i_v \sigma$ is a pairing of a one-chain and an $r$-chain which is mapped to a priori an $r-1$-chain but is a bit more general and the vector field is explicit:
\begin{align}
i: C^{(1)}(K,\, \mathbb{R}) \times C^{(r)}(K, \, \mathbb{R}) &\longrightarrow C^{(1)}(K, \, \mathbb{R})\times C^{(r-1)} (K,\, \mathbb{R}) \\
 (v,\, \sigma) &\longmapsto v \rfloor^K \eta(\sigma)
\end{align}
where $\eta(\sigma)$ is the $(r-1)$-chain that results from the contraction by taking a linear combination of faces of $\sigma^{(r)}$. There is also a sense in which the contraction is done in a ``local'' open set which is determined by the cell (or simplex) of highest dimension which contains both $v$ and $\sigma$. \\

\noindent
Also, the Whitney map applied to $i_v\sigma$ treats the vector field as a function coefficient i.e:
\begin{equation}
W (v \rfloor^K \eta(\sigma) ) = W (v) \rfloor^K \, W(\eta(\sigma)) = \varphi^{0}(W(v)) \wedge W(\eta(\sigma))
\end{equation}
where $\varphi^{{0}}$ deletes the constant form part of the Whitney element, 
\begin{equation}
\varphi^{0} (\omega dx) = \omega.
\end{equation}
The generalization of this operation when more than one contraction has been applied to a chain, like for the Jacobi identities below, is done in the obvious way. 

Having shown that the operation works in the cell, we return briefly to the consideration of chains, we alluded to them in the discussion of FIG. 3. There is a pairing of chains in $i_v \sigma$ which makes the map bilinear. It is clear that each simplex gives rise to a co-chain with support over all cells that have that simplex as a face. This is related to the notion of calculating in the ``local'' open set we alluded to. Otherwise, we are left with the consideration of constant vector fields, we will return to that in the discussion of the Lie derivative. We write the bi-linearity of the interior product as
\begin{equation}
i_v \sigma =  \sum_{i,j} i_{v_i} \sigma_j = \sum_j \sum_{\{ v_i | \, (v_i, \sigma_j) < \sigma_k^{(n)} \}} i_{v_i} \sigma_j.
\end{equation}
In words, given a  one simplex $v_i$ from the one-chain $v$,  and one term $\sigma_j$ from the chain $\sigma$, if they are both faces of a higher dimensional hyper-cube $\sigma_k ^{(n)}$, then the terms contributes. It is then straightforward to establish that the discrete interior product squared is zero.

Having introduced the interior product, we allow for geometry, but in doing so, the closure of the operations is on the product space of chains. We will establish that the other operations $\wedge$, \,$d$ and $\star$ can be modified in a way that makes the theory consistent topologically. Also, we will see that the Jacobi identities work exactly thus limiting the result of the chain operation to the product of two such spaces, while the intermediary steps involve three. 
So let us modify the discrete theory to accommodate the new operator $i_v$.
\subsection{Exterior derivative}
A priori it is already under control, since if $\omega$ is a Whitney form, we know how to calculate $d \omega$, it is a linear combination of $(r+1)$-degree Whitney elements using the co-boundary $d^K = \delta^K$ on the complex. However $di_v \sigma = d (v \rfloor^K \eta(\sigma))$ is problematic, we need to make a prescription for it since it is an object in the product space of chains. The outcome is that we can define it in a consistent way which also leaves the cohomology intact. To see this, go back to the 2D example and let
\begin{equation}
d^K (i_{[01]} [0123]) = d^K ([01] \rfloor^K ([0] + [1]) )= 2[01] \rfloor^K ([30] + [21])
\end{equation}
while mapping both sides to the continuum we get the correct result which is
\begin{equation}
d i_v ( (1-y) dx) = -2 (1-y) dy.
\end{equation}
The operation is well-defined in general, when applying $d$ after $i_v$, we only take partial derivatives with respect to the orthogonal directions to both $v^{(1)}$ and $\sigma^{(r)}$ and $r \geq 1$. So we can apply the Leibniz rule for functions in a combinatorial way by writing
\begin{equation}
\label{div}
d^K (i_v \sigma) = 2 v \rfloor^K \, d^K \eta(\sigma)
\end{equation}
where $d^K$ is the original co-boundary map, and thus, the homology of the theory in the product space is the same as that of the original theory and leads to the desirable result:
\begin{equation}
W (d^K i_v \sigma) = d W ( i_v \eta(\sigma)).
\end{equation}
To justify the formulae, we can give a proof, by taking the general Whitney form in a hyper-cube which was discussed above. So let 
\begin{equation}
W( \sigma^{(r)} ) = \Pi_{i=1}^{p} f_i^{-} \Pi_{j=p+1}^{n-r} f_j^{+} \, \bigwedge_{k=1}^{r} dx^{\mu_k}.
\end{equation}
Then, starting with the right hand side of \eqref{div}, we get
\begin{equation}
W ( 2 v \rfloor^K \, d^K \eta(\sigma)) = 2 (W (v)) \rfloor^K \,( \sum_{l} \frac{\partial}{\partial x^l}dx^{l} \wedge)  W (\eta(\sigma))
\end{equation}
the point is that when you move the derivatives $\frac{\partial}{\partial x^l}$ across to the left, you pick up all the terms since the vector has at least every $f_i$ factor that $\sigma$ has. There is a minor sign issue which as to be watched out for, since as stated above, we allow to contract with edges that are parallel to $\sigma$ and $\frac{\partial}{\partial x^l} f_l^{\pm} = \mp 1$.

Return to the co-homology. The $di_v$-property \eqref{div} can be used twice to show that
\begin{equation}
(d^K)^2 ( i_v \sigma) = v \rfloor^K \, (d^K)^2 \eta(\sigma) = 0
\end{equation}
since $(d^K)^2 = 0 $ on the space of chains. We had the property that 
\begin{equation}
(d^K)^2 C^r(K,\, \mathbb{R})=0,
\end{equation}
and we have deduced that 
\begin{equation}
(d^K)^2 i_v \, C^r(K,\, \mathbb{R})=0.
\end{equation}
The $di_v$-property is particularly convenient since it also helps to consider exact chains as well as closed chains. Let us consider an $(r+1)$-chain which is exact,
\begin{equation}
\label{exactf}
\eta^{(r+1)} = d^K \sigma^{(r)}.
\end{equation}
then, considering a one-chain representing the vector field $v$ and assume that we contracted an $(r+1)$-chain that gave $v \rfloor^K \sigma$. Then, immediately we get
\begin{equation}
v \rfloor^K \eta^{(r+1)} = \frac{1}{2} d^K (v \rfloor^K \sigma^{(r)}) = v \rfloor^K \, (d^K \sigma^{(r)}).  
\end{equation}
which is a rewriting of \eqref{exactf}
leading to the Homology groups
\begin{equation}
H^{(r)} (M, \, \mathbb{R})\cong B^{(r)}(M,\, \mathbb{R}) / Z^{(r)}(M,\, \mathbb{R}).
\end{equation}
The picture is the following: we had the link between the de Rham co-homology and the homology as a basis for discretising differential forms with the operator $d$. Now, we have extended the theory to the product space of chains in a way that preserves the content of the topological theory, and furthermore gives us access to the geometry. Whitney elements have a central role in the theory.
\subsection{The wedge product and the Hodge star}
The product is much the same, but it is worth pointing out how it co-exists with $i_v$ since we have to check that it acts as an anti-derivation on the space of chains i.e
\begin{equation}
\label{antidev}
i_v (\sigma^{(r)} \wedge^K \beta^{(p)} ) = (i_v \sigma^{(r)}) \wedge^K \beta^{(p)}  + (-1)^r \sigma^{(r)} \wedge^K i_v \beta^{(p)}.
\end{equation}
Again the criterion is given by the continuum analogue which is
\begin{equation}
W(i_v (\sigma^{(r)} \wedge^K \beta^{(p)} ) = W(i_v \sigma^{(r)}) \wedge^K \beta^{(p)})  + (-1)^r W( \sigma^{(r)} )\wedge^KW( i_v \beta^{(p)}) .
\end{equation}
But we have already proved that because the interior product is exact on the space of Whitney forms. So by using
\begin{equation}
i_{Wv} W = W i_v,
\end{equation}
we rely on the continuum identity and so \eqref{antidev}. As is known, the wedge gives rise to the cohomology ring, and that the discrete wedge (or cup product) is non-associative, or it is only up to a factor. By acting with $d$ on $\sigma \wedge^K \eta$, we invoke the Liebnitz rule which is exact only after integration (application of the de Rham map) in the original geometric discretisation. The point is that Whitney forms were only present as dummy integration variables which are compatible with $d$, while here we want exactness before integrating thus preserving the topological properties associated to $d$ and to $i_v$. Then, to take the wedge product of two ``contracted'' chains is done in the natural way (where $\otimes$ is the tensor product)
\begin{equation}
(\zeta \rfloor^K \eta(\sigma)) \wedge^K ( \beta \rfloor^K \eta(\alpha)) = (\zeta \otimes \beta) ( \eta(\sigma) \wedge^K  \eta(\alpha))
\end{equation}
again in a similar fashion as for $d$ you see that we are combining the homology groups in the same way as on the chain space:
\begin{equation}
H^{(r)} (K,\, \mathbb{R}) \wedge H^{(p)}(K,\, \mathbb{R}) \cong H^{(r+p)}(M,\, \mathbb{R}).
\end{equation}
as can be seen by noting that the term under brackets is the ordinary wedge operation in the chain space. \\

The Hodge star has two representatives in the discrete theory, one in the $K$ subdivision and one in the $L$ subdivision. We write:
\begin{equation}
\label{starex}
\star^K ( \beta \rfloor^K \eta(\sigma)) = \beta \rfloor^K ( \star^K \eta(\sigma)).
\end{equation}
Then, the associated properties of $\star^K, \, \star^L$ and $(d^K)^\dagger$ hold, since using \eqref{starex}, we get
\begin{equation}
(d^K)^\dagger = (\star^L d^K \star^K)(\star^L d^L \star^K) = \star^L (d^L)^2 \star^K = 0
\end{equation} 
where we used the fact that
\begin{equation}
\star^K \star^L = Id_L ,\, \star^L \star^K = Id_K.
\end{equation}
and in much the same way as for the sedge, we still have the Poincar\'e duality
\begin{equation}
H^{(r)}(M,\, \mathbb{R}) \cong H^{(n-r)} (M, \, \mathbb{R}).
\end{equation}
What is important at this point is that the operations $(\wedge, \, d, \, \star)$ are essentially topological and the function coefficient is passive or gives rise to a factor of two which does not modify the homology, a discussion of the co-homology of the theory in the product space is given in the appendix. Let us turn to the Lie derivative.
\section{Lie derivative and continuous symmetries}
With the operations $d$ and $i_v$ in place, it is a matter of consistency that the discrete Lie derivative, as given by the Cartan formulae 
\begin{equation}
L_v = i_v d + d i_v
\end{equation}
should have the following analogue:
\begin{equation}
L_v = i_v^K d^K + d^K i_v^K.
\end{equation}
We will find that we can go quite a long way with it and consider a discrete version of the Lie group.
As we saw, we consider the vector field as being represented by a one-cochain which is its dual differential form. The theory is seen as a theory of differential forms and so this choice is imposed to us. If we were to consider the vector fields in the dual complex as $\star^K v$ in the $L$, we would loose the common support that both the vector field an the form must have in order to carry through the construction of the exterior calculus. We start with examples, since we already have the operators.

 \underline{Example 1}: (see the Appendix), to investigate the construction, we start with the vector field (dual to) \eqref{vecteur} and the form \eqref{form}. These are the continuum analogies of \eqref{discretev}, \eqref{discretew}. The vector and the form will be considered on three faces, $[0176]$, $[1247]$ and $[0653]$ for this purpose. To each of those corresponds a volume element two-form, as given in section II (i.e a constant form such as $dx \wedge dz$). 
 Note that due to the orthogonality of the form with the vector chosen, we get $i_v v^b =0$. Next, evaluation of $dv^b$ in the discrete setting \eqref{divb} is obtained by means of applying the rules for the hyper-cubic complex given in section II. After application of the Whitney map the result (e.g \eqref{ivefour}) is obtained by application of the rule: \\



 This means that the transcription into an ordinary differential form is obtained by suppressing the form part (like a $dx$ or a $dx\wedge dy$) on the left hand side of the $\rfloor^K$ symbol. So by substituting the list of Whitney elements given in Appendix B, and the rule, we obtain \eqref{ivefour} which leads to the identification of the various terms in \eqref{liex1}. 
We have found the following:
\begin{equation}
\bar{W}( L_{\hat{v}} \hat{v}_{b} )= L_{W \hat{v}} W ( \hat{v}_b).
\end{equation}

We now move to a more delicate case, for which we will need to clarify the notion of the tangent space $T_{U}M$ of vector fields, where $U$ may be taken as an open set taken to be the cell. We wish to point out that we do not think of the tangent space $TM$ as a fiber bundle. The topology of the complex is too restrictive for that, open sets only overlap on a common boundary. However, the Whitney elements, because their support overlaps over boundaries guarantee as we will see, that we can consider properties of vector fields. \\



\underline{Example 2}: Take the vector $u$ to be parallel and in fact identical to $v^b$. In this case, one has to be careful since if we use the proposed formulae for the interior product, we get \eqref{internomix}, which is clearly wrong since there should be some terms with coefficient $v_{Z_1} v_{Z_2}$ as shown in the appendix. This is the point at which we introduce the rule above from which we get \eqref{intermix}. Considering \eqref{intermix}, keeping only the terms in $[0617]$, 
\begin{equation}
i_{\hat{u}} v_b = (v_{Z_2}^{2}- v_{Z_1} v_{Z_2}) [06]\rfloor^K ([0] + [6]) + (v_{Z_1}^{2}- v_{Z_1} v_{Z_2}) [17] \rfloor^K([1] + [7]),
\end{equation}
then,
\begin{align}
d i_u v_b &= 2(v_{Z_2}^2 - v_{Z_1}v_{Z_2}) [06] \rfloor^K ([10] + [76])+2(v_{Z_1}^2 - v_{Z_1}v_{Z_2} ) [17] \rfloor^K ( [01] + [67] ) \\
&=2 ( - (v_{Z_2}^2 - v_{Z_1} v_{Z_2}) (1-x) + ( v_{Z_1}^2 - v_{Z_1} v_{Z_2}) x ) dx \\
&=2 (v_{Z_1} - v_{Z_2} ) (v_{Z_2} (1-x) + v_{Z_1} x) dx 
\end{align}
as required to match the continuum expression. For the $i_ud$-term, the problem of support which lead us to modify the coefficients does not arise and we get
\begin{align}
i_u d v_b &=    (- v_{Z_2}^{2} [06] - v_{Z_1}v_{Z_2}[17] + v_{Z_1}^{2}[17] +v_{Z_1}v_{Z_2}[06]) \rfloor^K ([01] + [67] ) \\
&= (-v_{Z_2}^{2}(1-x) - v_{Z_1}v_{Z_2} x + v_{Z_1}^{2}x + v_{Z_1}v_{Z_2} (1-x)   )dx \\
&= (v_{Z_1} - v_{Z_2} ) (v_{Z_2} (1-x) + v_{Z_1} x) dx 
\end{align}

which gives the correct matching on $[0617]$, not forgetting the multiplicative factor of two specified by the rule for ``$di_v$''. This gives us the correct matching.

The combination of Example 1 and Example 2 cover all cases of a pair of a form and a vector field that can be expressed in the present framework (i.e parallel or orthogonal) . They constitute the two parts of the proof that the discretised Lie is the correct one. What we are describing here is a flow from the face $[0176]$ to the face $[0653]$ and check its vertical (z-component) at the edge $[06]$ which joins the two faces. The tangent vector to the integral curve is thus rotated by a right-angle as it propagates along its flow line. The component $v_x$ lives in $[0176]$, the component $v_y$ lives in $[0653]$ while the component $v_z$ lives in both. 

Evaluating components of the discrete Lie derivative as we have done above will be particularly useful when we turn to discuss the covariant derivative in the subsequent article. But before we address this issue, we ought to check how the Lie algebra property, and notably how the Jacobi identities translate to the lattice. This property gives us a lifting of the Lie algebra of vector fields (which we have yet to define) to the algebra of derivations acting on the space of forms. In the continuum we write:
\begin{equation}
\label{jaco}
\biglb[ L_v , \, L_u \bigrb] = L_{\biglb[v, \, u \bigrb]}. 
\end{equation}
Although we are in a position to evaluate the left-hand side of \eqref{jaco}, we need to make a prescription for the complex based bracket $[v, \, u]$, and in fact we know that in the infinite dimensional representation of vector fields which we adopt, the continuum commutator is clear:
\begin{equation}
\label{infliealg}
\biglb[u, \, v \bigrb] = (u^{i} \partial_i v^j - v^i \partial_i u^j) \partial_j.
\end{equation}
Or conversely, we need to show that starting with $\biglb[L_v, \, L_u\bigrb]$, the vector field which is the argument of the Lie derivative on the right-hand side is indeed the discrete analogue of $[v,\, u]$ given in \eqref{infliealg}. \\

\noindent
Given two one-chains $\sigma$ and $\eta$ on a $2$D complex, we introduce their bracket 
\begin{align}
\biglb[\, . \, , \, . \, \bigrb]_K : C^1 (K)\, \times \, C^1 (K) &\longrightarrow C^{2}(K) \times C^1 (K) \\
               \left([\sigma], \, [\eta]\right)  &\longmapsto \biglb[\sigma, \, \eta \bigrb]_K
\end{align}
where in components, assuming summation over the indices, we are given $\sigma = \sigma_{ij} [ij]$ and $ \eta = \eta_{ij} [ij]$, then,
\begin{equation}
\label{liebracket}
\biglb[v, \, u \bigrb]_K = (\sigma_{ij}\eta_{kl} -\eta_{ij}\sigma_{kl})  [ijkl] \rfloor^K [ij].\\
\end{equation}

\noindent
Example: Again, much is achieved by an example. Take the face $[0671]$, and using the previous example, we take:
\begin{align}
\hat{u} &= v_{X_1} [01] + v_{X_2} [67] \\
\hat{v} &= v_{Z_2} [06] + v_{Z_1} [17]
\end{align}
The continuum analogies are
\begin{align}
W \hat{u} &= v_{X_1} (1-z) dx + v_{X_2} zdx \\
W \hat{v} &= v_{Z_2} (1-x) dz + v_{Z_1} xdz
\end{align}
Application of \eqref{infliealg}, leads to
\begin{equation}
\biglb[ W u, \, W v\bigrb] =  \biglb(v_{X_1} (1-z)  + v_{X_2} z \bigrb)\biglb(v_{Z_1}- v_{Z_2} \bigrb)dz +  \biglb(v_{Z_2} (1-x)  + v_{Z_1} x\bigrb)\biglb(v_{X_2}-v_{X_1}  \bigrb)dx
\end{equation}
It is straightforward to reproduce this formulae on the lattice by using the prescription. 

The construction can be clarified further by considering the example of $so(3)$, the rotations in 3D. We will setup the Lie algebra and show that it is well-defined on the complex. Second, we will explain how, although the Jacobi identities are found to hold, the lifting to the group is only approximate which is because integral curves are approximated by Whitney elements. Consider the $so(3)$ generators and the form $\omega$ respectively given by
\begin{align}
X &= y \partial_z - z \partial_y, \\
Y &= z \partial_x - x \partial_z, \\
Z &= y \partial_x - x \partial_y, \\
\omega &= W ([65]) = (1-x)z dy.
\end{align}
A direct calculation in the continuum leads to
\begin{equation}
L_{[X, \, Y]} W([65]) = [L_X,\, L_Y] W([65]) = - z(1-x)dx - yz dy
\end{equation}
which are the continuum Jacobi identities for the Lie. 
We now proceed to analyse the discrete version of the Jacobi identities by constructing the complex based version of this calculation. We will use the rules we have introduced, notably the $d i_v$-rule, the rule for immersion and the rule for products of coefficients given below. The central feature is that of closure, although we use a product space, we can recover the exact expressions by simply identifying:
\begin{equation}
1 \wedge \phi = \phi,
\end{equation}
after applying the latter rule. In the 3D coordinates available, the generators are exactly represented by:
\begin{equation}
\hat{Y} = ([17] + [24]) - ([67] + [54]),\; \hat{X} = ([35] + [24]) - ([65] + [74]).  
\end{equation}
because a look at the table of Whitney elements in the appendix, will help to see that these are the dual one forms of the vector fields after mapping them to the chains to the continuum. The bracket we proposed was guided by requiring 1) The closure of the algebra under the bracket, 2) The functional form of $[X, \, Y]$, 3) The Jacobi identities.
\begin{equation}
[X, \, [Y, Z]] + [Y, \, [Z, X]] + [Z,\, [X,Y]] = 0 
\end{equation}
and at the level of the Lie derivative it reads:
\begin{equation}
L_{[X, \, Y]} = L_X L_Y - L_Y L_X,
\end{equation}
we will derive this relation below for the complex based theory.
This means that we seek a complex based bracket operation that maps to the continuum one by a Lie algebra morphism. With the basis of forms we have, we can already set what the answer is:
\begin{equation}
\label{brac}
[\hat{X}, \, \hat{Y}]^K = A  ([ W(X), \, W(Y)])  
\end{equation}
and this is satisfactory because it is an exact relation and not an approximation as:
\begin{equation}
W [ \hat{X}, \, \hat{Y}]^K = Z 
\end{equation}
the continuum $Z$, which means that $W$ is a Lie algebra morphism. So this would be enough to establish 2), and certainly proves the existence of the discretised bracket.

What we have is that for a Lie algebra $\bf{g}$ in the infinite dimensional representation of vector fields. The map $W$ is a Lie algebra morphism if and only if, the collection of one form dual to the generators of $\bf{g}$ can be written as a linear combination of Whitney elements.

The bracket in \eqref{brac} is an implicit definition making use of the continuum one. It is satisfactory as we were able to give an exact analogue of the vector fields $X,\, Y, \, Z$.
We proceed to the calculation of the Jacobi identities for the Lie, and first of all we need the list of edges that are mutually parallel, so that we will know when to apply the rule called ``Tangent space of vector fields'': 
\begin{equation}
L_1 = \{[17],\, [06],\, [35],\, [24] \}, \, 
L_2 = \{[67], \, [54],\, [32], \, [01] \}, \,
L_3 = \{ [65], \, [74],\, [12],\, [03]\}.
\end{equation}
The detailed calculation is done in Appendix C, since all the rules we introduced play a part, we displayed in curly brackets the associated continuum calculation for the purpose of illustration. The picture should then be clear, the algebra of vector fields is local and is exactly captured. The Jacobi identities for the algebra
\begin{equation}
ad ([x, \, y]) = [ ad (x), \, ad(y)]
\end{equation}
which are satisfied here, have a ``local'' analogue in the Lie group, consider the vector fields $X_x$, $X_y$, satisfying
\begin{equation}
[X_x, \, X_y] = X_{[x,\,y]}
\end{equation}
and the flow of $X_x$ is given by
\begin{equation}
\Phi_x (t,\, a) = a \exp(tx).
\end{equation}
But the latter flow is only approximate since it cannot be captured in the space of Whitney elements (for example a vector field rotating in the $x,y$ plane.
The limitation is easy to understand, owing to the exactness of the vector fields in the theory, the algebra is exact. However the space of co-chains as being given by the Whitney elements does not lead to the correct integral curves and the associated relations, we only obtain an approximation, which is for the flow of a vector field $\Phi_x$ and its discrete counterpart $\hat{\Phi}_x$ is controlled by correction terms of order $\Phi_x - W (\hat{\Phi}_x )$.
In conclusion to the part on the Lie algebra and lie groups, we gather the results. A Lie group action on the space of chains and associated Lie algebra of vector fields are such that: 

\noindent
1) The Lie algebra is exactly captured if the generators are elements of the space of Whitney elements. \\
2) The associated Lie group $G$ is exactly captured if the integral curves of the vector fields leave the space of Whitney elements invariant. Otherwise it is approximated by the Whitney forms.

We now address the central issue of implementing the technique we have just presented and expose some numerical results.

\section{Convergence of the model and numerical implementation.}

 We have seen that the results are exact on the space we have introduced, however the Jacobi identity for the Lie is only satisfied in the local picture, flows in the group are only approximate in general. Overall, this means that we have good control over the convergence. 

An issue related to convergence is that of refining the triangulation (which is here made of squares). A discussion of this in the context of simplicial categories \cite{thomas} is worth considering. In that picture as here, one works with chain complexes $C_.$ and their dual co-chain complexes and it is argued that they are to be preferred to homology. Due to the Whitehead theorem they are invariant for connected spaces. That is, we take subdivisions of the original complex and one can relate the associated chain spaces. In turn, given the chain space and the dual cochain space, one can define the various mappings we have introduced above. 
\begin{figure}[H]
\label{fig:T}
{\par\centering \includegraphics{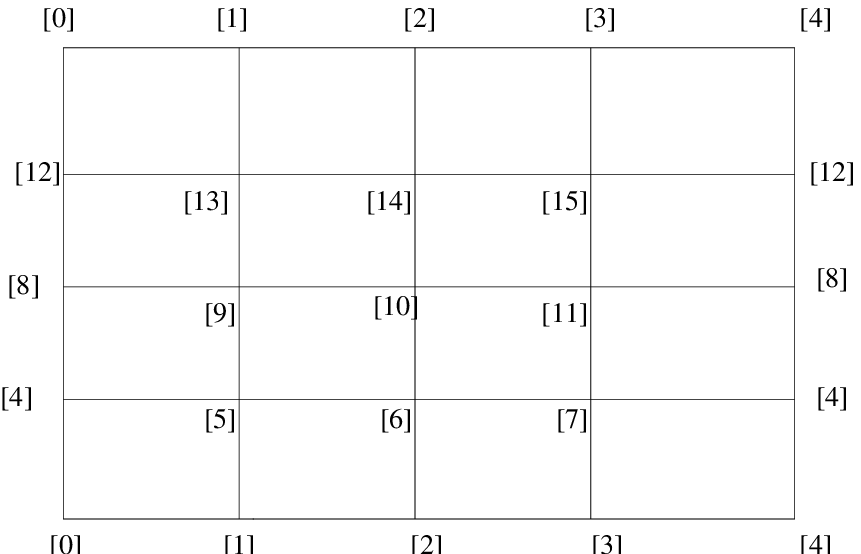} \par}
\end{figure}
In the present perspective, we insist on the fact that refining the triangulation is the only mean by which one can obtain a better approximation to the various forms which we always take to be Whitney elements.

Next, we discuss the example of a one-form field on the torus, satisfying the equation 
\begin{equation}
\label{fluiddy}
(L_v - \eta \triangle) \omega = 0 
\end{equation}
which is relevant to fluid dynamics \cite{arnold}, and is a first step toward setting up the Navier-Stokes equation in this formalism. We start by tackling the problem analytically and then numerically. The square ``triangulation'' of the torus is shown above:

Now, the vector field is constant, taken along the $x$-axis in the plane of the page $v = v_x \partial_x$.
We seek a one-form solution as a linear combination of Whitney elements. The Lie derivative is then:
\begin{equation}
L_v \omega = (y_+ - y_-) v_x dy 
\end{equation}
The notation we use for components, is such that it can be applied to every $2$D cell. The sign and $i$ or $j$ indices denote a specific edge. For example in $[56,10,9]$, $x_+$ is the component on the edge $[9,10]$ and $y_-$ is the component on the edge $[59]$. Similarly when a component outside the cell is needed, we take for example $y_{+,j}$ to be the $[10,14]$ component. Next is the Laplacian term:
\begin{align}
d d^{\dagger} \omega &= ( - ( x_+ - x_{+,-i} + y_{-, j} - y_-) + (x_{+,i} - x_+ + y_{+, j} - y_+)) [x+] \\
\label{five}
&+ ((x_+ - x_{+,-i}+ y_{-,j} - y_-) - (x_- - x_{-,-i} + y_- - y_{-,-j}))[y-] \\
\label{seven}
&+ ((x_{+,i}-x_+ + y_{+,j} - y_+) - (x_{-,i}- x_- + y_+ - y_{+, -j})       )[y+] \\
&+ ((x_{-, i}- x_- + y_+ - y_{+,-j})-(x_- - x_{-,-i}+ y_- - y_{-,-j})) [x-]  
\end{align}
and
\begin{align}
d^{\dagger} d \omega &=(( y_{+,j} - x_{+,j}  - y_{-,j} + x_+) - (y_+ - x_+ - y_- + x_-)        ) [x+] \\
\label{six}
&+  ((y_+ - x_+ - y_- + x_-)-(y_- - x_{+,-i} - y_{-,-i} + x_{-,-i})   )[y-] \\
\label{eight}
&+ ((y_{+,i} - x_{+,i} - y_+ + x_{-,i}) - (y_+ - x_+ - y_- + x_-) )[y+] \\
&+ (( y_+ - x_+ - y_- + x_+) -(y_{+,-j} - x_- - y_{-,-j} + x_{-,-i})) [x-]
\end{align}
The first solution is the constant zero-mode solution. Then, $L_v$ is zero because the $y$-edges have equal coefficients across each ribbon. Which means that:
\begin{align}
y_{-,-i} &= y_- = y_+ = y_{+,i}= Y \\
y_{-,j} &= y_{+,j}= Y_+ \\        
y_{-,-j} &= y_{+,-j}= Y_- 
\end{align}

In this case, equation \eqref{fluiddy} reduces to:
\begin{equation}
\triangle \omega = 0.
\end{equation}

To get zero we need either each component be zero or each pair of coefficients (e.g the two $[x+]$ components) to cancel out. Set all $x$ to be zero and we are left only with:
\begin{equation}
\triangle \omega = (Y_+ - 2 Y + Y_-) [y-] + (Y_+ - 2 Y + Y_-) [y+]
\end{equation}
Our equation is finally
\begin{equation}
\label{first}
Y_+ - 2 Y + Y_- = 0
\end{equation}
This gives us a constant form $\omega = C dy$ as an exact solution.
Let us think about \eqref{first}. After all this is an approximation to the second derivative:

\begin{equation}
(\partial_x)^2 \omega_y = Y_+ - 2 Y + Y_-
\end{equation}
by using a finite difference rule for the derivative:
\begin{equation}
D_x \omega = \frac{1}{h} ( \omega(x+h) - \omega(x))
\end{equation}
which is correct to order $h$. Since $[y-]$ has the same coefficient as $[y+]$ makes the associated form one with a constant coefficient and so having Whitney elements presents no benefit from the usual finite difference approach in terms of approximation in this case.



In order to move to a more general example, we consider again a solution in $y$-components, but this time we let $y$ increase as we move from one $y$-ring to the next one, that is in every cell we let $y_+ - y_- = \theta$. Then,
\begin{equation}
L_v \omega = (y_+ - y_-) v_x dy = \theta v_x dy
\end{equation}
In order to cancel the Lie term, we require as before that all $x$-coefficients be zero and so
\begin{align}
\label{firstone}
y_+ - y_- &= y_{+,j} - y_{-,j} \\
\label{secondone}
y_+ - y_- &= y_{+,-j} - y_{-,-j} \\
\label{thirdone}
y_+ - y_- &= y_{+,j} - y_{-,j} \\
\label{fourthone}
y_+ - y_- &= y_{+,-j} - y_{-,-j}  
\end{align}
Equate \eqref{five} and \eqref{seven}. We find that they have equal coefficients after noting that using \eqref{firstone} we can rewrite \eqref{five} as
\begin{equation}
y_{-,j} - y_- - y_- + y_{-,-j} = y_{+,j} - y_+ - y_- + y_{-,-j} 
\end{equation}
similarly for \eqref{seven}. Next, equate \eqref{six} and \eqref{eight}.
\begin{align}
y_+ - y_- - ( y_- - y_{-,-i}) &= y_{+,i} -y_+ -y_+ +y_- \\
y_+ - y_- &= \frac{1}{3} (y_{+,i} - y_{-,-i})  
\end{align}
We are now in a position to write down a new solution.
Let $y$ be constant along each ring parallel to the $y$-axis. Also, we have seen that
\begin{equation}
y_+ - y_- = \theta
\end{equation}
If we enforce this for all squares we have a local (in the sense of a single simplex) matching of Whitney elements which is fine. However, if we are to look at boundary conditions, particularly for closed surfaces like the torus, we will need some more prescription to guarantee closure of the field. To this end we consider 
\begin{equation}
\phi = a e^{i\frac{x}{b}}dy
\end{equation}
where $a$ and $b$ are constants to be determined.
Then, in the interior of a given square, we have
\begin{equation}
\phi = a ( 1 + i \frac{x}{b})dy + O( x^2)
\end{equation}
and letting $x= ix$, leads to the usual expression, after we enforce the following better ansatz:
\begin{align}
\hat{\phi} &= \phi( l_1) ( 1 - \frac{x}{b} ) + \phi(l_2) (  \frac{x}{b} ) \\
\phi (l_1) &= \int_{| l_1|} a e^{i\frac{x}{b}}dy = a e^{i \frac{x(l_1)}{b}}.
\end{align}
Moreover if $b= h$ the edge length of a small square, then we can match the function with the Whitney map applied to a chain field $\hat{\phi}$.
We the have the matching
\begin{equation}
\phi = W ( \hat{\phi}) + O(x^2)
\end{equation}
To move with the analogy one step further, note that
one retains a bijection between the function $e^{ix}$ and $ix$. i.e, the function $\phi$ oscillates over a large number of squares:
\begin{equation}
\| y_+ - y_- \| << 2 \pi.
\end{equation}
To guarantee closure, let $N_x$ be the number of edges along a circle of the torus (parallel to the x-axis), then require closure (recall $y_+ - y_- = \theta$ a constant):
\begin{equation}
(y_+ - y_-) N_x = 2 k \pi. \; \;\;\;\; \;  \{k = 0, \, 1, \, 2, \, \ldots\}
\end{equation}
Then, $k=0$ leads to the zero mode we already got, the rest give the $k$-th mode solution. Let us do the continuum calculation:
The change of variable $x \mapsto i x$ leads to $dx \mapsto i dx,\; \frac{\partial}{\partial x} \mapsto -i \frac{\partial}{\partial x},\; v_x \mapsto -i v_x$. Apply this to the solution:
\begin{align}
L_v \phi &= v_x \frac{1}{b} \frac{\partial}{\partial ix} e^{i \frac{x}{b}} dy = v_x \frac{a}{b} e^{i \frac{x}{b}} dy \\
\star d \star d \phi &= \frac{a}{b^2} e^{i \frac{x}{b}} (dy)
\end{align}
This solves the equation provided $\frac{v_x}{b}= \eta$. In order to move closer our analogy between the continuum and the complex we need the following, evaluate:
\begin{align}
d \ln \phi \, dy &= \frac{ \phi '}{\phi} dx \wedge dy \\
\star d \ln \phi \, dy &= \frac{ \phi '}{\phi} \\
i_v d \ln \phi \, dy &= v_x\frac{ \phi '}{\phi}\, dy
\end{align}
Then,
\begin{align}
d\star d \ln \phi \, dy &= \frac{ \phi ''\phi - \phi' \phi'}{\phi^2}dx \\
\star d \star d \ln \phi\, dy &=\frac{ \phi ''\phi - \phi' \phi'}{\phi^2}(-dy)= \frac{\phi''}{\phi} (-dy) + \bigglb(\frac{\phi'}{\phi}\biggrb)^2 dy.
\end{align}
Note that according to the discrete theory we get:
\begin{equation}
\label{origdif}
\frac{\phi'}{\phi} = \frac{e^{i x_{k+1}} -e^{i x_k}}{e^{ix_k}} = i ( x_{k+1} - x_k) + O((x_{k+1} - x_k)^2)
\end{equation}
Using the modulus,
\begin{align}
\| \bigglb(\frac{\phi'}{\phi}\biggrb)^2  \| \sim \| \frac{\phi'}{\phi} \|^2 \sim \epsilon^2
\end{align}
Therefore,
\begin{align}
\triangle \ln \phi - L_v \ln \phi &= -\frac{\phi''}{\phi} - v_x \frac{\phi'}{\phi} + O(\epsilon)^2 =\frac{1}{\phi} \bigglb[ - \phi'' - v_x \phi' \biggrb] + O ( \epsilon^2) \\
&= \frac{1}{\phi} \bigglb[ \triangle \phi - L_v \phi \biggrb] + O ( \epsilon^2) 
\end{align}
In the discrete approach this is written as
\begin{equation}
A^1(\ln \phi) = \sum_j \int_{[j_1 j_2]}  \frac{2k i\pi x_j}{a}dy  [j_1 j_2] = \sum_j  \frac{2k i\pi x_j}{a} [j_1 j_2]
\end{equation}
Then apply the discrete version of the operator 
\begin{equation}
L_v - \triangle  
\end{equation}
 and solve the matrix equation. Next, with a varying vector field, the solution is modified, that is the rate of change of the solution starts varying. This does not modify the calculation above with the minor exception that now
\begin{equation}
\| \frac{\phi'}{\phi} \| = \| \frac{ f(x_{k+1}) e^{i x_{k+1}} - f(x_k) e^{i x_k}}{f(x_k) e^{ix_k}}  \| < \sup \bigglb[ \|\frac{ f(x_{k+1})}{f( x_k)}\|, \, \, 1 \biggrb] O (\epsilon^2) \sim O ( \epsilon^2),
\end{equation}
which leads to the same equation as before with a variable $v_x$.

Regarding the inclusion of modes, we have studied the function $\ln \phi$ where in principle we could include all the modes (along the x-circles):
\begin{equation}
\phi = \sum_k e^{2 \pi ik x} f_k ( x, \, y)dy 
\end{equation}
Let
\begin{equation}
\chi = f_k e^{\kappa x}.
\end{equation}
Then we will get
\begin{align}
(\triangle - L_v) \ln (\chi) &= \frac{1}{\chi}( - \chi'' - v_x \chi') \\
&= \frac{1}{\chi} \bigglb[ - (f_k''- 2 \kappa f_k' - \kappa^2 f_k ) - v_x (f_k \kappa + f_k')\biggrb]e^{\kappa x}
\end{align}
Substituting functions,  note the $x^2$ term for $f_k$. 
\begin{align}
f_k&= a+ bx + c x^2, \\ 
v_x&= v_0 + v_x x,
\end{align}
leads to the following equations for the constant term and the $x$-term:
\begin{align}
a \kappa^2 +(2b - v_0 a) \kappa - 2c - v_0 b &= 0 \\
b \kappa^2 +(4c-v_0 b -v_x a) \kappa - v_x b -2 v_0 &=0
\end{align}
These equations are to be solved for $a, \, b$ and $c$ in principle. We can then substitute back the solution into our trial solution:
\begin{equation}
\chi = (a + bx + c x^2) e^{ \kappa x} 
\end{equation}
thus
\begin{equation}
\chi ' = (  \kappa ( a + bx + c x^2) + b + 2cx )e^{\kappa x}
\end{equation}
The $x$ term in this expression is then
\begin{equation}
(b + 2c + a \kappa)
\end{equation}
The approximation is then given by:
\begin{align}
\hat{\chi} &=  (<f>_0 + <f>_1 x)e^{\kappa x} \\
           &= <f>_0  + (<f>_0 \kappa+ <f>_1)x  + O (x^2)
\end{align}
So
\begin{equation}
\chi' = (<f>_0 \kappa+ <f>_1)(x_{n+1} - x_{n} ) + O (\epsilon^2)
\end{equation}
which is to be compared with \eqref{origdif}. The observation being that the theory captures the linear term. We leave here such considerations of matching the continuum solution.

To gather our findings we note that the key features are the use of the logarithm and the selection of the mode by specifying the relevant boundary condition. The $k$-th mode is given by specifying the constant $C$ which is a normalisation and setting:
\begin{align}
\phi^+ &= C \\
\phi^- &= C + 2 \pi k 
\end{align}

Numerical tests have been done on the torus we have displayed above (with a few extra squares only), and although we have already discussed the convergence, we would like to explain the numerical procedure. It consists of using a singular value decomposition (SVD) algorithm \cite{reci}. 
\vspace{0.3cm}
{\par\centering \resizebox*{3in}{3in}{\includegraphics{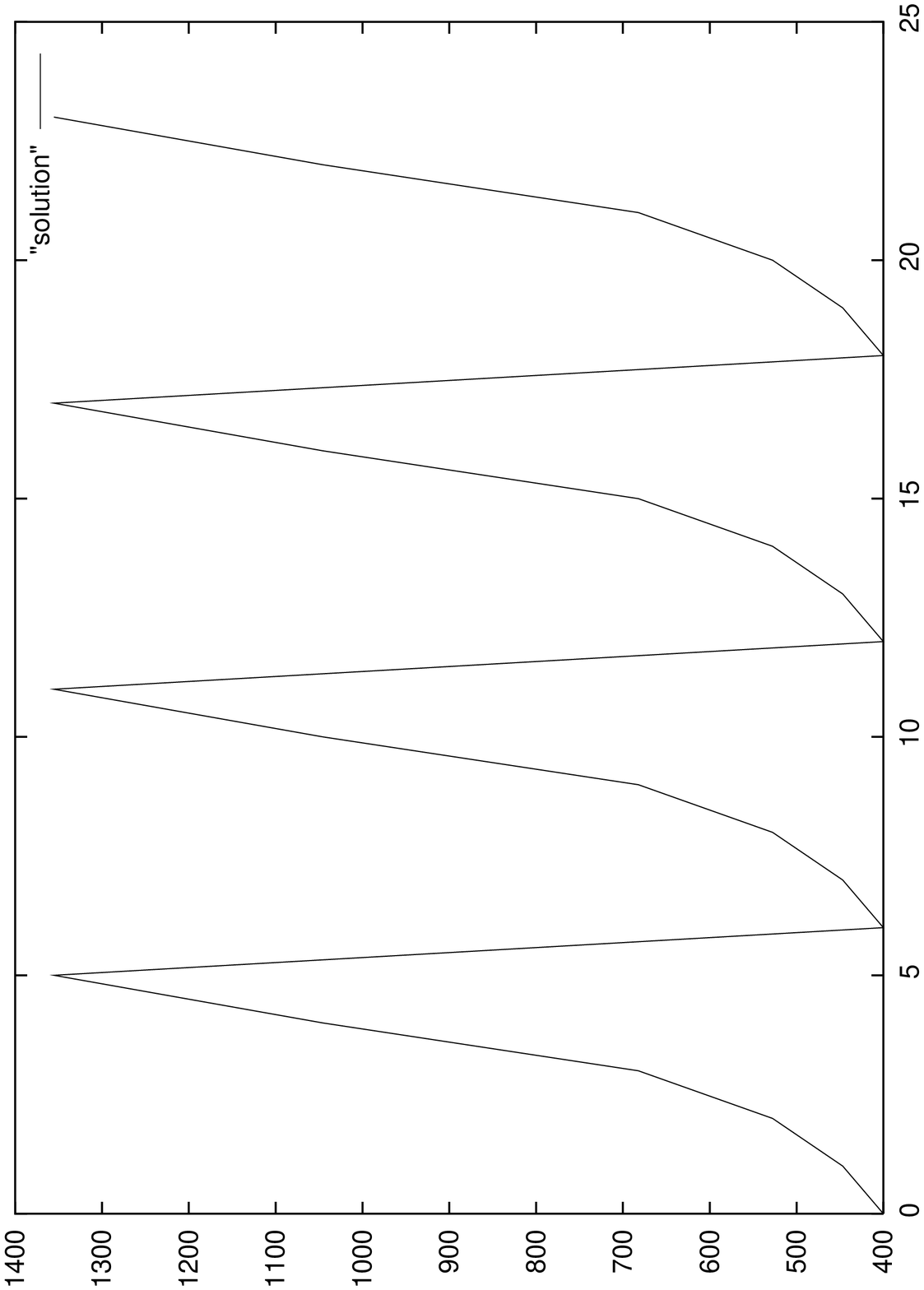}} \par}
\vspace{0.3cm}

One form field solution with values in the range $[400,\, 1400]$ over circles parallel to the $x$-axis. The solution is constant over $y$-circles, e.g $[1],[7],[13],[19]$ are the vertices of one such circle. \\

Two key advantages to this method are first its well-known stability and second that it solves a matrix equation by minimizing the square norm $\chi^2$ below, so in some cases were the present theory cannot be solved exactly as a matrix equation, the algorithm gives us a good approximation. The SVD is used to solve
\begin{equation}
A x = b
\end{equation}
and leads to the following decomposition (the $w_j$ are the singular values):
\begin{equation}
A^{-1} = V . [diag \,\frac{1}{w_j}] U^T.
\end{equation}
The solution is then obtained by minimizing 
\begin{equation}
\chi^2 = \| Ax - b \|^2.
\end{equation}
We also note that the formal errors due to the $\chi^2$ fit, let $V_{(1)}\ldots V_{(M)}$ be the columns of $V$ then the boundaries of the confidence ellipsoids are given by:
\begin{equation}
\triangle \chi^2 = w_1^2 (V_{(1)} . \delta a)^2 + \ldots + w_{M}^2 (V_{(M)} . \delta a)^2.
\end{equation} 
In the case of the graph above, we took the following vector field:
\begin{align}
v_x &= 4, \,\,\,\,\,\, \forall x \in [0, \,3[, \, \forall y; \\ 
v_x &= -2, \,\, \forall x \in [3, 6[, \forall y; \\
v_y &= 0 , \, \, \forall \,(x, \, y).
\end{align}
Note that the change in the vector field results in a change of concavity in the graph as the boundary conditions are enforced on the solution. Here $C = 400$ and we identify $\phi^+ = 400$ with $\phi^- = 1400$. In this example the matrix $A$ is $96 \times 96$. For a constant vector field, we just get a straight line in each section. The inclusion of a non-zero mode is achieved by modifying the matrix in a way that take into account the quotienting which lead to the principal value of $i\phi$. The modes are then treated one at a time, by specifying the appropriate matrix.

\section{Conclusion}

We have proposed a new discretisation scheme for differential geometry which is deeply related to the corresponding continuum theory. The first step which is central, was to accommodate an interior product. This required the introduction of a product space. Then, preserving the de Rham complex motivated us to consider a ring of operators composed of $i_v$, $d$ and $\star$, and the introduction of a special rule for $d i_v$, for which a prescription was required. The Lie derivative, as an element of the ring, required us to modify the coefficient of the chain so as to accommodate the case in which the vector field varies over the cell as expressed by the sum of Whitney elements. The bracket was also introduced for vector fields in the infinite dimensional representation, for which the geometrical interpretation is crucial. Also, the Jacobi identities for the Lie is exact, but one bears in mind that in general the flows are approximated by Whitney forms. We also have some numerical implementation for the torus and a hold on convergence formally. We are now in a position to introduce metrics in this scheme which we proceed to do in the article II.
\section{Appendix}

\subsection{Hyper-cubic Whitney elements}

For the purpose of this article we need the list of Whitney elements given in the table.

\begin{table}[here]
\begin{center}
\begin{tabular}{|c|c|c|}
\hline
Zero-cochains & One-cochains & Two-cochains \\
\hline
$\mu_0 = (1-x)(1-y)(1-z)$ & $W([01]) = dx(1-y)(1-z)$ & $W([0123]) = (1-z) dx \wedge dy$ \\
$\mu_1 = x(1-y)(1-x)$ & $W([03]) = dy(1-x)(1-z)$ & $W([1247]) = xdy \wedge dz$ \\
$\mu_2 = xy(1-z)$ & $W([06]) = dz (1-x)(1-y)$ & $W([0356]) = (1-x)dx\wedge dz$\\
$\mu_3 = y(1-x) (1-z)$ & $W([17]) = x(1-y)dz$ & $W([0167]) = (1-y)dx \wedge dz$\\
$\mu_4 = xyz$ & $W([12]) = x (1-z) dy$ & $W([6745]) = z dx \wedge dy$\\
$\mu_5 =(1-x)y z$ & $W([24]) = xydz$ & $W([3245]) = ydx \wedge dz$ \\
$\mu_6 = (1-x)(1-y) z$ & $W([32]) = y(1-z)dx$ & $$          \\ 
$\mu_7 = x(1-y) z$ & $W([35]) = y(1-x)dz$ & $$\\
$$ & $W([65]) = z(1-x)dy$ & $$ \\
$$ & $W([67]) = (1-y)zdx$ & $$ \\
$$ & $W([74]) = xzdy$ & $$ \\
$$ & $W([54]) = (1-z)zdx$ & $$ \\

\hline
\end{tabular}
\end{center}
\end{table}

\subsection{Relation to the de Rham complex}

Let us discuss the relation of the setup to the de Rham complex, which is extracted from differential forms under the exterior derivative $d$. Having shown that on the extended space (after application of $i_v$), $(d^K)^2 = 0$ is a priori not sufficient. But note that we have apparently lost one important feature, which is the de Rham theorem \cite{nakhara}, which states that the map $\Lambda$, defined as
\begin{equation}
\Lambda: H_r (M) \times H^r (M) \longrightarrow \mathbb{R}
\end{equation}
is bilinear and non-degenerate. This map could be expressed as
\begin{equation}
(\sigma, \, W(\eta)) = \int_\sigma W(\eta). 
\end{equation}
However, with the construction introduced above, we seem to have lost the non-degeneracy property of $\Lambda$. To see this, let $\zeta^r \in C^r ( K, \, \mathbb{R})$, and let $\eta^{(r+1)}\in C^{r+1}(K ,\, \mathbb{R})$. Then using the rule for immersion, we get
\begin{equation}
\int_{\sigma^{r}} \varphi^{0}(W ( v)) \wedge W (i_{\hat{v}} \eta)
\end{equation}
to be compared with 
\begin{equation}
\int_{\sigma^{r}} W (\zeta).
\end{equation}
Clearly, the space of forms spanned by products of chains is larger, but we have shown, in the proof of the Hyper-cubic rule that the action of $d^K$ on the larger space has some functorial properties which link it with $d$, we found that,
\begin{equation}
\bar{W} d^K i_v = d \bar{W} i_v
\end{equation} 
and so we may put the various steps of that calculation under the integral symbol. It means that we have a larger space of forms and that the co-homology is ``intact''. This is another virtue that we extract from the $di_v$-rule.

\begin{equation}
\begin{array}{ccc}
\Omega^{r}(M, \, \mathbb{R})                 &
\stackrel{id}{\longrightarrow} &
\Omega^{r}(M, \, \mathbb{R})                             \\
\Big\uparrow{\Lambda}             & & 
\Big\uparrow{\Lambda}   \\
C^{p}\left(K, \, \mathbb{R} \right)&  
\stackrel{id}{\longrightarrow}&
C\left(K, \, \mathbb{R}\right) \otimes C^{p} \left(K, \, \mathbb{R} \right)
\end{array}
\end{equation}

A comment on the diagram: we wrote $\Lambda$ where we could have written $W$, but we stress that we are using the inner product of a space of chains with the space of differential forms. The vertical arrow on the right relates two larger spaces than that on the left. The diagram being commutative entails the compatibility of the present scheme with the original geometric discretisation. One could say that the introduction of geometry does not modify topology.

In terms of the homology we got,
\begin{equation}
H^{r} (K\times K, \, \mathbb{R}, \, \bar{d}^K ) \cong H^{r} ( K, \, \mathbb{R}, \, d^K ), 
\end{equation}
while,
\begin{align}
B^{r}(K, \, \mathbb{R}, \, d^K)  &< B^{r} (K \times K, \, \mathbb{R},\, \bar{d}^K), \\
Z^{r}(K, \, \mathbb{R}, \, d^K)  &< Z^{r} (K \times K, \, \mathbb{R},\, \bar{d}^K).
\end{align}

Hence, the extended space of chains consisting of elements of the form specified by the rule for immersion form a larger subspace of the space of forms under the modified Whitney map than does the ordinary chains under the Whitney map. The extended space can be embedded into the product space 
which has the same topology.

\subsection{Examples}

\underline{Example 1}: Let
\begin{align}
\label{vecteur}
v &= v_{X_1} ( 1 - z ) dx + v_{X_2} z dx + v_{Y_1} ( 1 - z)dy + v_{Y_2} z dy \\
\label{form}
v^{b} &= v_{Z_2} ( 1 - x ) dz  + v_{Z_2} ( 1 - y ) dz + v_{Z_1} x dz + v_{Z_3} ydz
\end{align}
Note that 
\begin{equation}
i_v v^b = 0.
\end{equation}
Next,
\begin{equation}
d v^b = - v_{Z_2} dx \wedge dz - v_{Z_2} dy \wedge dz + v_{Z_1} dx \wedge dz + v_{Z_3} dy \wedge dz, 
\end{equation}
\begin{align}
\label{liex1}
L_v v^b= i_v d v^b &= - v_{Z_2}( v_{X_1}(1-z) + v_{X_2}z)dz 
- v_{Z_2} (v_{Y_1}(1-z) + v_{Y_2} z )dz \\
&+ v_{Z_1} (v_{X_1}(1-z) + v_{X_2} z )dz +v_{Z_3} (v_{Y_1}(1-z) + v_{Y_2} z )dz
\end{align}
Lattice calculation:
\begin{align}
\label{discretev}
\hat{v} &= v_{X_1} [01] + v_{X_2} [67] + v_{Y_1} [03] + v_{Y_2}[65] \\
\label{discretew}
\hat{v_b} &= v_{Z_2} [06] + v_{Z_1}[17] + v_{Z_3} [35] 
\end{align}
Again, see the discussion of the interior product, 
\begin{equation}
i_{\hat{v}} \hat{v_b} = 0 
\end{equation}
Then,
\begin{equation}
\label{divb}
d \hat{v_b}= (v_{Z_1} - v_{Z_2}) [6017] + (v_{Z_2} - v_{Z_3}) [0653] +v_{Z_1} [1724] + v_{Z_3}[3542] 
\end{equation}
and
\begin{align}
\label{iveone}
i_{\hat{v}} ( v_{Z_1} - v_{Z_2}) [6017] &=(v_{Z_1} - v_{Z_2})(  v_{X_1} ( [01] \rfloor^K [06] + [01] \rfloor^K [17] ) +  v_{X_2} ( [67] \rfloor^K [06] + [67]\rfloor^K [17]) )
 \\
i_{\hat{v}}(v_{Z_2} - v_{Z_3} )[0653]&= (v_{Z_2} - v_{Z_3}) ( v_{Y_1} ( [03] \rfloor^K [06] + [03] \rfloor^K [35] ) + v_{Y_2} ( [65] \rfloor^K [06] + [65]\rfloor^K [35]) )
\end{align}
the continuum analogue of which is
obtained by identifying the sum of Whitney elements with a constant form,
\begin{align}
\label{ivefour}
\bar{W}i_{\hat{v}} ( v_{Z_1} - v_{Z_2}) [6017]&= (v_{Z_1} - v_{Z_2}) ( v_{X_1}((1-z)dx \rfloor  dz ) +   v_{X_2}(zdx \rfloor dz )) \\
\bar{W} i_{\hat{v}} (v_{Z_2}- v_{Z_3} ) [0653]&=-   (v_{Z_2} - v_{Z_3}) v_{Y_1} (((1-z)dy \rfloor dz ) -  v_{Y_2}(zdy \rfloor dz) )  
\end{align}
Which matches the continuum expression after use of the continuum identification, simply by deleting the $\rfloor$ sign and the constant differential which is part of the coefficient form.

\underline{Example 2}:
Next, try with $u$ which is parallel to $v_b$. Then,
\begin{equation}
\label{iuvd}
i_{u} v_{b} = ( v_{Z_2} (1-x) + v_{Z_1}x )^2 + ( v_{Z_2} ( 1 - y) + v_{Z_3} y )^2
\end{equation}
and,
\begin{equation}
\label{diuvb}
d i_u v_b = 2 ( v_{Z_1}- v_{Z_2} )(v_{Z_2} ( 1-x) + v_{Z_1} x ) dx + 2 (v_{Z_3} - v_{Z_2} ) ( v_{Z_2} ( 1 - y) + v_{Z_3} y ) dy 
\end{equation}
Next $dv_b$ was given above and,
\begin{equation}
\label{iudvb}
i_u d v_b =  (v_{Z_2} (1-x) + v_{Z_1} x) (v_{Z_2} - v_{Z_1})dx +  ( v_{Z_2}( 1- y) + v_{Z_3} y) (v_{Z_2} - v_{Z_3} )dy
\end{equation}
and finally,
\begin{equation}
\label{luvb}
L_u v_b = ( v_{Z_1} - v_{Z_2} ) (  v_{Z_2} ( 1-x) +  v_{Z_1}x  )dx + (v_{Z_3} - v_{Z_2}) ( v_{Z_2} ( 1-y) +  v_{Z_2} y  )dy 
\end{equation}
Lattice calculation:
\begin{equation}
\label{u}
\hat{u} = v_{Z_2} [06] + v_{Z_1} [17] + v_{Z_3}[35] =v_b
\end{equation}
Then,
\begin{equation}
\label{internomix}
i_{\hat{u}} v_b = v_{Z_2}^{2} [06]\rfloor^K ([0] + [6]) + v_{Z_1}^{2} [17] \rfloor^K([1] + [7]) + v_{Z_3}^{2} [35] \rfloor^K ( [3] + [5] ) 
\end{equation}
in order to reproduce the cross terms we set:
\begin{align}
\label{intermix}
i_{\hat{u}} v_b &=  (v_{Z_2}^{2} [06]+ v_{Z_1} v_{Z_2}[17])\rfloor^K ([0] + [6]) +( v_{Z_1}^{2} [17]+ v_{Z_1}v_{Z_2}[06]) \rfloor^K([1] + [7]) \\
&+( v_{Z_3}^{2} [35]+ v_{Z_3}v_{Z_2} [06]) \rfloor^K ( [3] + [5] )
\end{align}
and
\begin{equation}
d^K i_{\hat{u}} v_b =(v_{Z_2}^{2} [06]+ v_{Z_1} v_{Z_2}[17])\rfloor^K([10] + [76]) + ( v_{Z_1}^{2} [17]+ v_{Z_1}v_{Z_2}[06]) \rfloor^K([01] + [67])
\end{equation}
Next, we already calculated $d v_b$, so that 
after retaining the components in $[0653]$ and $[0176]$,
\begin{align}
2i_{\hat{u}} d v_b &=- v_{Z_2}(  v_{Z_2} [06]  + v_{Z_1} [17] ) \rfloor^K ( [01] + [67] ) + v_{Z_2} ( v_{Z_2}[06] - v_{Z_3} [35] ) \rfloor^K( [03] + [65]) \\
&+ v_{Z_1} ( v_{Z_1} [17] + v_{Z_2} [06] ) \rfloor^K( [01] + [67]
) + v_{Z_3} ( (v_{Z_3} [35]  - v_{Z_2} [06]) \rfloor^K ( [03] + [65])  
\end{align}

Gathering the terms, we get
\begin{align}
i_{\hat{u}} d v_b &= ( -v_{Z_2}^{2} [06] - v_{Z_1}v_{Z_2}[17] + v_{Z_1}^{2}[17] +v_{Z_1}v_{Z_2}[06]) \rfloor^K ([01] + [67] ) \\
&+ ( v_{Z_2}^{2}[06] - v_{Z_2}v_{Z_3}[35] + v_{Z_3}^{2}[35] - v_{Z_3}v_{Z_2} [06])\rfloor^K([03] + [65] ) 
\end{align}


\underline{Example 3}: The complex based calculation is as follows:
\begin{align}
i_X [65] &= -([65] + [74]) \rfloor ([6] + [5]) \\
   \{ &= - z^2 (1-x)\}  \\
i_Y [65] &= 0 \\
d^K i_X [65] &= -2 ([65] + [74]) \rfloor ([06] + [76] + [35] + [45]) \\
\{ &= -2z(1-x) dz + z^2 dx \} \\
d^K [65] &= [6574] + [6530] \\
\{ &= (1-x) dz \wedge dy - z dx \wedge dy \} \\
i_X d^K [65] &= -([65] + [74]) \rfloor ([67] + [54])  + ([65] + [74]) \rfloor ([06] + [35]) + ([35] + [24]) \rfloor ([65] + [03]) \\
\{ &= y(1-x) dy + z (1-x) dz - z^2 dx \} \\           
i_Y d^K [65] &=  (-[67]-[54]) \rfloor ([65] + [74])- ([17] + [24]) \rfloor ([03] + [65])  \\
\{ &= - x(1-x) dy - z^2 dy \}
\end{align}
So,
\begin{align}
L_X [65] &= -([65] + [74])\rfloor( [06] + [35]) + ([35] + [24]) \rfloor ([65] + [03]) \\
&\{= y(1-x) dy  - z(1-x) dz            \} \\
L_Y [65] &= i_Y d^K [65] 
\end{align}
Next,
\begin{align}
i_Y L_X [65] &= ([65] + [74]) \otimes ([17] + [24]) \rfloor ([0] + [6] + [3] + [5]) \\
&+ ([65]+ [74]) \otimes (-[67] - [54]) \rfloor (- [7] - [6] - [4] - [5] )  \\
\{ &= 2z(1-x)x  - xz (1-x)  \}   \\
d^K i_Y L_X [65] &= 2 ([65] + [74] ) \otimes ([17] + [24]) \rfloor ([10] + [76] + [23] + [45] ) \\
&+ 2([65] + [74] ) \otimes ( -[67] - [54] ) \rfloor (-[17] - [06] - [24] - [35] ) \\
\{ &=  2z(1-x) dx + 3z^2dz - 2 zx dx + 2 (1-x) x dz - z(1-x) x +  xzdx - 3 z^2 dz - x(1-x) dz \}   \end{align}
\begin{align}
d^K i_X d^K [65] &= ([65] + [74] ) \rfloor ([6701] + [5423])  - ([65] + [74]) \rfloor ( [0617] + [3542]) \\
&- ([35] + [24]) \rfloor ([6574] + [0312]) \\
\{ &= -z dx \wedge dz - 2 z dz \wedge dx - y dx \wedge dy \} \\
i_Y d^K i_X d^K [65] &=  2 ([65] + [74]) \otimes ([17]+[24]) \rfloor(-[01] - [67] + [54] + [32]) \\
&+ 2 ([65] + [74]) \otimes (-[67] - [54]) \rfloor (-[35] - [42]- [06] - [17] ) \\
&+  (-[35] - [24]) \otimes (-[67] -[54]) \rfloor ([65] + [74] + [03] + [12]) \\
\{ &= - z^2 dz - xzdx + 2 z^2 dz + 2 xz dx - yz dy \} \\
i_Y i_X d^K [65] &= ([65] + [74]) \otimes (-[67] - [45]) \rfloor (-[6]-[7] - [5] - [4] ) \\
&- ([65] + [74]) \otimes ([17] + [24]) \rfloor ([0] + [6] + [3] + [5]) \\
\{ &= - xz(1-x) - z^3 \} \\
d^K i_Y i_X d^K [65] &=  ([65] + [74]) \otimes (-[67] - [45]) \rfloor (-[06]-[17] - [35] - [24] )  \\ 
&- ([65] + [74]) \otimes ([17] + [24]) \rfloor ([10] + [76] + [23] + [45])  \\
\{ &= -z (1-x) dx + xzdx - 3 z^2 dz - x(1-x) dz \}
\end{align}
then
\begin{align}
L_Y L_X [65] 
&= - z (1-x) dx + zxdx - 3 z^2dz - x(1-x) dz  \\
&+ 2z (1-x) dx + 3 z^2 dz - 2zxdx + 2 (1-x) x dz \\
&- z^2 dz - xzdx + 2 z^2 dz + 2xzdx - yz dy  \\
&= z(1-x) dx + x(1-x) dz+ z^2 dz  - yzdy 
\end{align}
Exact matching is also found in the calculation of $L_Y L_X [65]$.
\begin{acknowledgments}
V de B would like to thank his supervisor James C. Sexton, and Mike Peardon for help with the numerical issues. The work of V de B was supported by the Higher Education Authority of Ireland through the IITAC project. 
\end{acknowledgments}

\end{document}